\documentclass[aps,pre,twocolumn,showpacs]{revtex4}
\usepackage{amssymb}
\usepackage{epsfig}

\begin{document}

\title{\Large \bf Phase diagram and critical behavior of the square-lattice Ising model with competing nearest- and next-nearest-neighbor interactions}

\author{Junqi Yin and D. P. Landau}

\affiliation{Center for Simulational Physics, University of
Georgia, Athens, Georgia 30602, USA\\
}

\date{\today}

\begin{abstract}
Using the parallel tempering algorithm and GPU accelerated
techniques, we have performed large-scale Monte Carlo simulations of
the Ising model on a square lattice with antiferromagnetic
(repulsive) nearest-neighbor(NN) and next-nearest-neighbor(NNN)
interactions of the same strength and subject to a uniform magnetic
field. Both transitions from the $(2\times1)$ and row-shifted
$(2\times2)$ ordered phases to the paramagnetic phase are
continuous. From our data analysis, reentrance behavior of the
$(2\times1)$ critical line and a bicritical point which separates the
two ordered phases at T=0 are confirmed. Based on the critical
exponents we obtained along the phase boundary, Suzuki's weak
universality seems to hold.
\end{abstract}


\maketitle

\section{Introduction}

For the nearest-neighbor (NN) Ising antiferromagnet on the square
lattice in a uniform magnetic field, the low temperature ordered
phase is separated from the paramagnetic phase by a simple, 2nd
order phase boundary.  (Within the context of the lattice gas model
this system could be described as having repulsive NN-coupling and
forming a $c(2 \times 2)$ ordered state.)  With the addition of
repulsive (antiferromagnetic) next-nearest-neighbor (NNN)
interactions the situation becomes more complicated.  Early Monte
Carlo simulations suggested that a single, super-antiferromagnetic,
or $(2 \times 1)$, phase existed, separated from the paramagnetic
phase by a single phase boundary\cite{Lan71,Lan80,Bin80}.  A
degenerate, row-shifted $(2 \times 2)$ state was also predicted at
zero temperature.  (See Fig.~\ref{f1} for a schematic representation of these states.)  On the other hand, symmetry arguments based on
Landau theory\cite{Dom78} predict the order-disorder transitions of
$(2\times1)$ and $(2\times2)$ structures belong to XY model with
cubic anisotropy.

The original motivation of this study was to investigate the
possibility of XY-like behavior of  the Ising spins on the square
lattice, since there is numerical evidence\cite{Lan83} that Ising
antiferromaget with attractive NNN interaction on the triangular
lattice has an XY-like intermediate state between the low
temperature ordered state and high temperature disordered state. In
fact, the present model has already been studied by many authors
using different approaches. An early Monte Carlo study\cite{Bin80}
comprehensively showed the phase diagrams for several different
interaction ratios (R) of NNN to NN interaction. But due to the
deficiencies of computer resources at that time, for the $R=1$ case,
a disordered region was missed between the two ordered phases which
was pointed out in a later interfacial free energy
study\cite{Slo83}. Meanwhile, transfer matrix
studies\cite{Kas83,Ama83} found reentrant behavior for the
$(2\times1)$ transition lines. While a study using the cluster
variation method\cite{Lop93,Lop94} concluded that for a range of R
(0.5$\sim$1.2) the system undergoes a first order transition, a
recent Monte Carlo study\cite{Mal06} using a variant of the
Wang-Landau method\cite{wls} focused on the $R=1$ case without
external field and found the phase transition is of second order.
For external field $H=4$ (in the unit of NN interaction constant)
the two ordered phases, namely $(2\times1)$ and row-shifted
$(2\times2)$, are degenerate at zero temperature so it is tempting
to think that the cubic anisotropy would be zero for this field and
that there could be a Kosterlitz-Thouless transition.

In this paper, we carefully study the location of phase boundaries
and the critical behavior for the case $R=1$. In Sec.~\ref{s1} the
model and relevant methods and analysis techniques are reviewed. Our
results are presented in Sec.~\ref{s2}, along with finite size
scaling analyses, and we summarize and conclude in Sec.\ref{s3}.

\section{Model and method}\label{s1}

\subsection{The model}
The Ising model with NNN interaction is described by the Hamiltonian
\begin{equation}
{\cal
H}=J_{NN}\sum_{<i,j>_{NN}}\sigma_i\sigma_j+J_{NNN}\sum_{<i,j>_{NNN}}\sigma_i\sigma_j+H\sum
\sigma_i,
\end{equation}
where $\sigma_i, \sigma_j= \pm1$, $J_{NN}$ and $J_{NNN}$ are NN and
NNN interaction constants, respectively, H is an external magnetic
field, and the sums in the first two terms run over indicated pairs of neighbors on a
square lattice with periodic boundary conditions.  Both $J_{NN}$ and $J_{NNN}$ are positive (antiferromagnetic) and the ratio $R=J_{NNN}/J_{NN}$.

\begin{figure}
\includegraphics[width=3.25in]{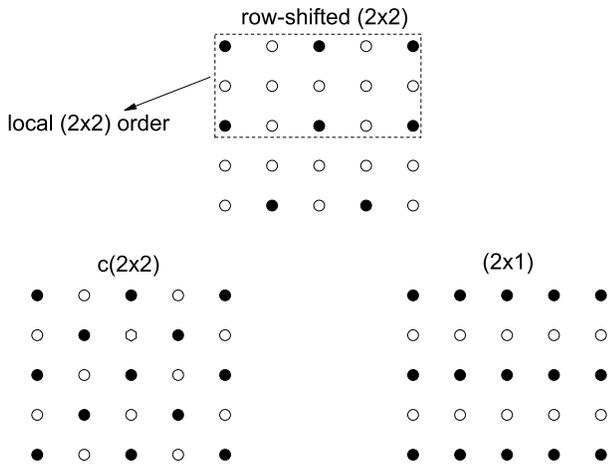}
\caption{Schematic plots of $c(2 \times 2)$, $(2 \times 1)$ and
row-shifted $(2 \times 2)$ ordered structures within the context of
the lattice gas model.  (In magnetic language, filled circles correspond to up spins and empty circles correspond to down spins.)} \label{f1}
\end{figure}

For the $R=1$ case, the ground states would be the $(2\times1)$
state, also known as super-antiferromagnetic state, in small
magnetic fields; and at higher fields it would be a  row-shifted
$(2\times2)$ state, which differs from the $(2\times2)$ state in the
sense that the antiferromagnetic chains in the former state can
slide freely without energy cost. See Fig.~\ref{f1}. Locally, the
structure may appear to be $(2\times2)$, but for large enough
lattices the equilibrium structure always shows row shifting. As a
result, such a row-shifted $(2\times2)$ state is highly degenerate,
and the antiferromagnetic sublattice exhibits only one dimensional
long range order. In terms of the sublattice magnetizations
\begin{equation}
M_\lambda = \frac{4}{N}\sum_{i\in\lambda}\sigma_i, \quad
\lambda=1,2,3,4
\end{equation}
we can define two components of the order parameter for the
$(2\times1)$ state
\begin{equation}
M^{a}=[M_1+M_2-(M_3+M_4)]/4,
\end{equation}
\begin{equation}
M^{b}=[M_1+M_4-(M_2+M_3)]/4,
\end{equation}
with a computationally convenient root-mean-square order parameter
\begin{equation}
M^{rms}=\sqrt{(M^{a})^2+(M^{b})^2}.
\end{equation}
Since $M^{rms}$ would have a limiting value of $\frac{1}{2}$ for the
row-shifted $(2\times2)$ state and be zero for the disordered
state, it can also be used as an order parameter for the row-shifted
$(2\times2)$ state.

Other observables, such as the finite lattice ordering
susceptibility $\chi$ and fourth-order cumulant $U$, are defined in
terms of the order parameter $M^{rms}$ as
\begin{equation}
\chi=\frac{N}{T}\left[<(M^{rms})^2>-<M^{rms}>^2\right]
\end{equation}
\begin{equation}
U=1-\frac{<(M^{rms})^4>}{3<(M^{rms})^2>^2}
\end{equation}
where N is the total number of spins and T is the simulation
temperature. In some cases, the true ordering susceptibility
$\chi^{+}$, which is $\frac{N}{T}<(M^{rms})^2>$, is used to
eliminate simulation errors resulting from $<M^{rms}>$, where the
order parameter is known to be zero for the infinite lattice.

\subsection{Simulation methods}
For small lattice sizes, Wang-Landau sampling~\cite{wls} was used to obtain a quick overview of the thermodynamic behavior of our model.  A two-dimensional random walk in energy and magnetization space was performed so that the density of states $g(E,M)$ could be used to determine all thermodynamic quantities (derived from the partition function) for
any value of temperatures and external field.  Consequently, "freezing" problems are avoided at extremely
low temperatures.  This allowed us to determine the ``interesting'' regions of field-temperature space; however, it quickly became apparent that, because of subtle finite size effects, quite large lattices would be needed.  Unfortunately, as L increases, the number of entries of histogram explodes as $L^4$ and it proved to be more efficient to use parallel tempering instead.

Since a large portion of interesting phase boundary is at relatively
low temperatures and many local energy minima exist which makes the
relaxation time rather long, the parallel tempering
method\cite{Swd86,Hukushima96} is a good choice for simulating our
model. The basic idea is to expand the low temperature phase space
by introducing configurations from the high temperatures. So, many
replicas at different temperatures are simulated simultaneously, and
after every fixed number of Monte Carlo steps, a swap trial is
performed with a Metropolis-like probability which satisfies the
detailed balance condition. The transition probability from a
configuration $X_m$ simulated at temperature $\beta_m$ to a
configuration $X_n$ simulated at temperature $\beta_n$ would be
\begin{equation}
W(X_m,\beta_m|X_n,\beta_n)=min[1,exp(-\triangle)],
\end{equation}
\begin{equation}
\triangle=(\beta_n-\beta_m)({\cal H}_m-{\cal H}_n).
\end{equation}
We chose the temperatures for the replicas to be in a geometric
progression\cite{Kof04}, which would make acceptance rates relatively
constant among neighboring temperature pairs, and the total number
of temperatures was chosen to make the average acceptance rate
above $20\%$.

The multiplicative, congruential random number generator RANECU was used \cite{James90,Ecuyer88}, and some results were also obtained using the Mersenne Twister \cite{twister} for comparison.  No difference was observed to within the error bars.

Typically, data from $10^6$ to $10^7$ MCS  were kept for each run and 3 to 6 independent runs are taken to calculate standard statistical error bars. For parallel termpering, the swap trial was attempted after every MCS.
In all the plots of data and analysis shown in following sections, if error bars are not shown they are always smaller than the size of the symbols.

In general, such replica exchange not only applies to the
temperature set, but also can apply to any other sets of fields,
such as the external magnetic field. Following the same argument,
one can get the transition probability from $\{X_m, H_m\}$ to
$\{X_n,H_n\}$
\begin{equation}
W(X_m,H_m|X_n,H_n)=min[1,exp(-\triangle)],
\end{equation}
\begin{equation}
\triangle=\beta(H_n-H_m)(M_m-M_n).
\end{equation}
where $M_m, M_n$ are the uniform magnetizations of replica m and n,
respectively.

\subsection{Finite-size scaling analysis}
To extract critical exponents from the data, we performed
finite-size scaling analyses along the transition lines. Since the
maximum slope of the fourth-order cumulant $U$ follows\cite{Fer91}
\begin{equation}\label{eq1}
(\frac{dU}{dK})_{max}=a^{\prime}L^{\frac{1}{\nu}}(1+b^{\prime}L^{-\omega}),
\end{equation}
where $K=\frac{J_{NN}}{k_BT}$,  the correlation length exponent
$\nu$ can be estimated directly.

With the exponent $\nu$ and critical temperature $T_c$ at hand, the
critical exponent $\beta$ and $\gamma$ can be extracted from the
data collapsing of the finite-size scaling forms,
\begin{equation}
M=L^{-\frac{\beta}{\nu}}\overline{X}(tL^{\frac{1}{\nu}})
\end{equation}

\begin{equation}
\chi T=L^{\frac{\gamma}{\nu}}\overline{Y}(tL^{\frac{1}{\nu}})
\end{equation}

\noindent where $t=|1-\frac{T}{T_c}|$, and $\overline{X}$ and $\overline{Y}$
are universal functions whose analytical forms are not known. One
can also estimate the exponent $\alpha$ from the relation of the
peak position with lattice size for the specific heat
\begin{equation}\label{eq2}
C_{max}=cL^{\frac{\alpha}{\nu}}+C_0
\end{equation}
where $C_0$ is the "background" contribution. In some cases when the
appropriate paths, i.e. which are perpendicular to the phase
boundary, are ones of constant temperatures, then the critical
behavior would be expressed in terms of reduced field
$h=|1-\frac{H}{H_c}|$, and all the above analysis still applies.

\subsection{GPU acceleration}
General purpose computing on graphics processing units(GPU) attracts
steadily increasing interest in simulational physics\cite{Yan07,
And08, Pre09}, since the computational power of recent GPU exceeds
that of a central processing unit(CPU) by orders of magnitude. The
advantage continues to grow as the performance of GPU's doubles
every year. Recently, a GPU accelerated Monte Carlo simulation of
Ising models\cite{Pre09} was performed. Compared to traditional CPU
calculations, the speedup was about 60 times.

The idea behind the implementation in Ref \cite{Pre09} can be easily
extended to our model, and the parallel tempering algorithm is
naturally realized. Initially, all the replica are loaded to the
global memory of the GPU. For each replica, the entire lattice is
divided into four sublattices, then spins in the same sublattice can
be updated simultaneously by the GPU using a Metropolis scheme, and
the swap of configurations of replica pairs can also be achieved in
parallel.

On the GeForce GTX285 graphics unit, our code runs about 10 times faster than it
does on the 32 CPUs of IBM p655 cluster using Message Passing
Interface(MPI) for parallel computation.

\begin{figure}
\includegraphics[width=0.95\columnwidth]{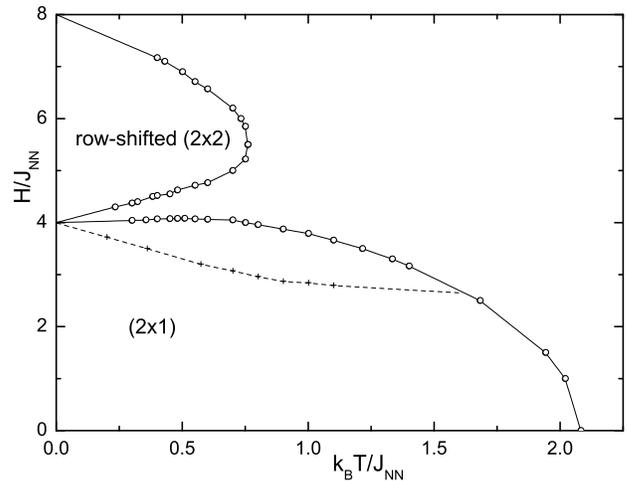}
\caption{The phase diagram for the Ising square lattice with
antiferromagnetic nearest- and next-nearest-neighbor interactions in
a magnetic field for $R=1$. Open circles and pluses denote
simulation results. The solid lines are second order transition
lines, while the dashed line indicates the short range ordering
line.} \label{f2}
\end{figure}

\section{Results and Discussion}\label{s2}
\subsection{Phase diagram and short range ordering}
From the ground state analysis, with zero or low field the ordered
state would be the superantiferromagnetic, or $(2\times1)$,
structure.  As the external field increases to $4<H/J_{NN}<8$, more
spins align in the opposite field direction, and the ordered
structure would be the row-shifted $(2\times2)$. With even stronger
fields, the system becomes paramagnetic.  In the region near
$H/J_{NN}=4$ a mixture of $(2\times1)$ and row-shifted $(2\times2)$
is visible.

\begin{figure}
\includegraphics[width=0.95\columnwidth]{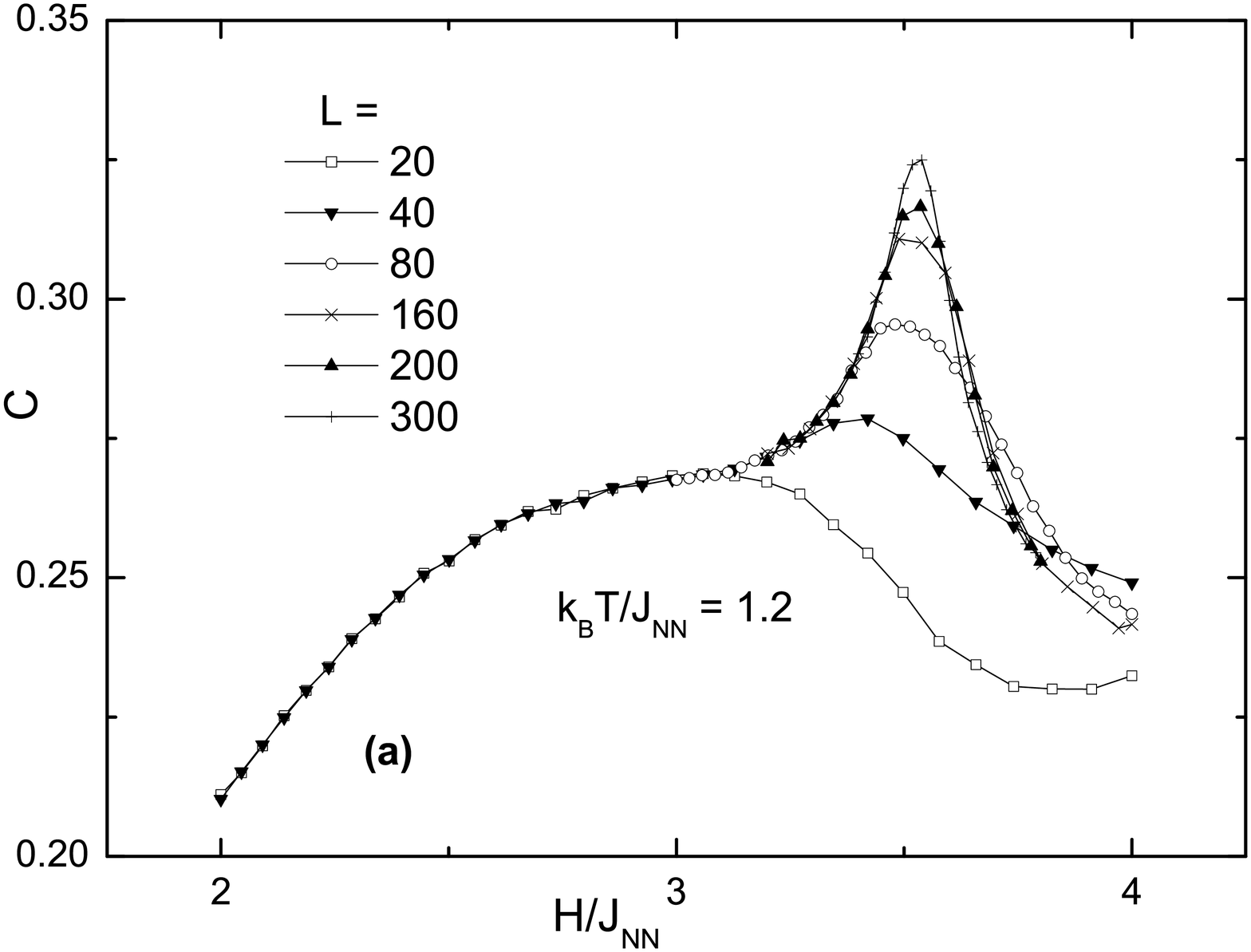}
\vspace{0.5cm}\\
\includegraphics[width=0.95\columnwidth]{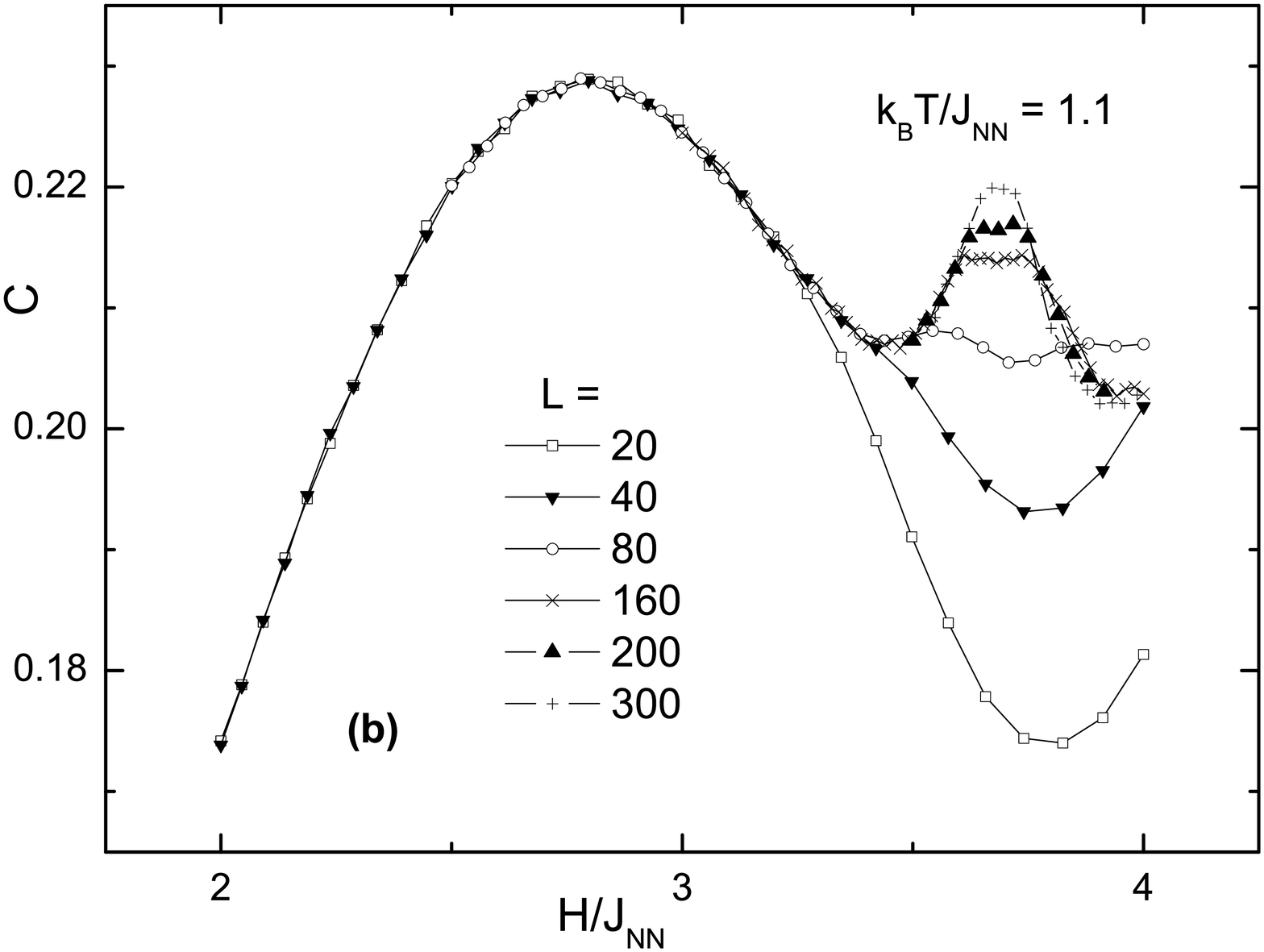}
\caption{Variation of the specific heat $C$ versus field $H$ with
lattice sizes $L=20, 40, 80, 160, 200, 300$ for paths of constant:
(a)$k_BT/J_{NN}=1.2$ and (b)$k_BT/J_{NN}=1.1$.} \label{f3}
\end{figure}

For finite temperatures, we found that the fourth-order cumulant is
always a good quantity to use to locate the transition points, while the
data for other quantities, such as the specific heat or susceptibility,
may look "strange" due to the effect of neighboring critical
points. To help the reader understand the observed thermodynamic properties, the final phase diagram for $R=1$ is plotted in Fig.~\ref{f2}:  The
 solid lines are the phase boundaries, all of which are
continuous. The dashed line inside the $(2\times1)$ ordered phase
indicates a ``short range ordering'' line, which was located from
the peak position of the specific heat. As shown in Fig.~\ref{f3},
for paths of constant temperature $k_BT/J_{NN}=1.1$ and $1.2$, there
are two peaks, and the one that increases with lattice size
corresponds to the critical point.

An indication of the complexity of the finite size behavior is
clearly seen in the bottom portion of Fig.~\ref{f3} in which the
small lattices actually have minima in the specific heat for field
values that eventually show phase transitions for sufficiently large
systems.  The round-shaped size-independent peak is due to the short
range ordering of the $(2\times1)$ ``clusters'' of different
orientation from the ordered background. No corresponding
behavior was observed from the susceptibility or the fourth-order
cumulant.

\begin{figure}
\includegraphics[width=0.95\columnwidth]{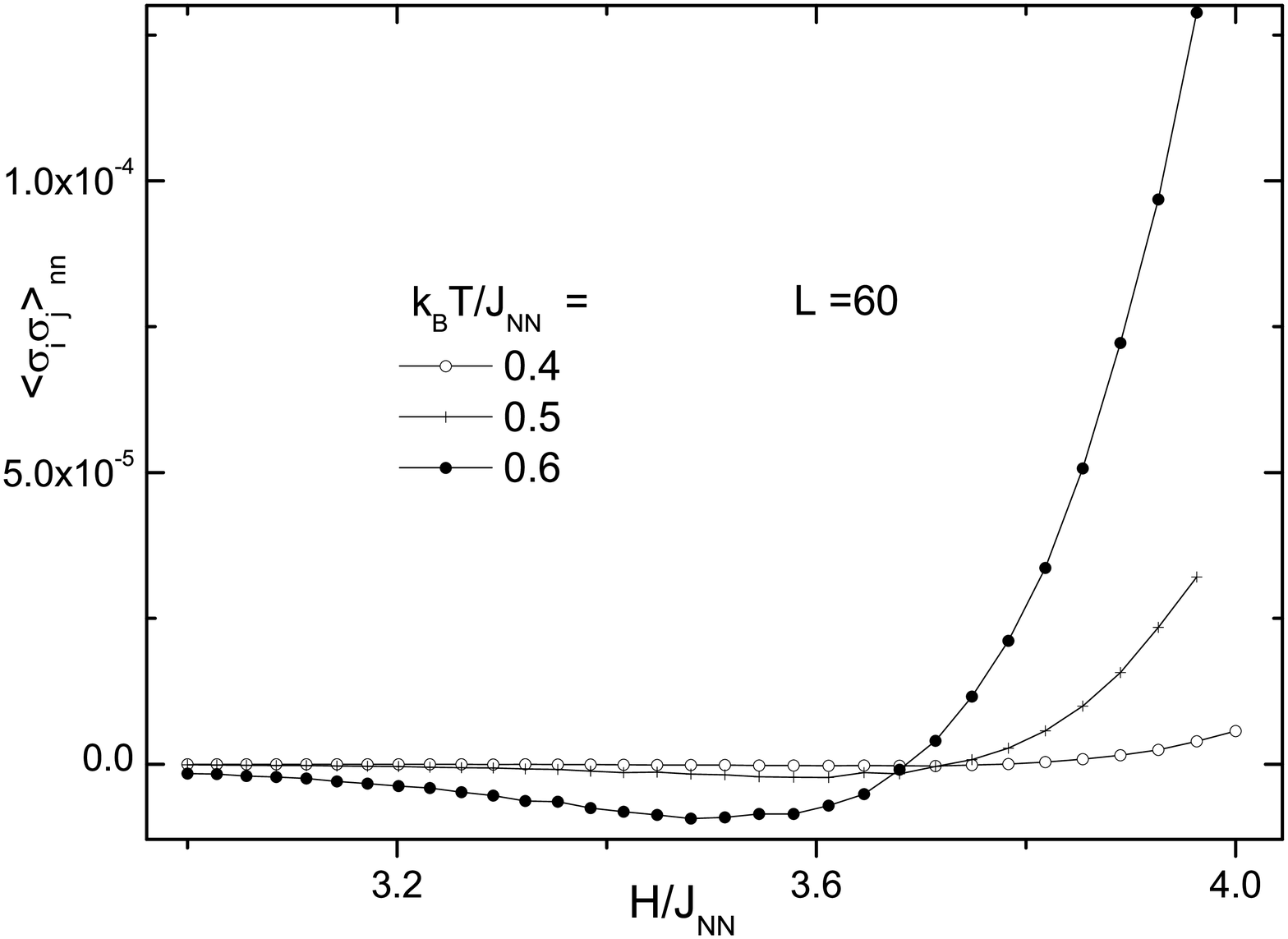}
\vspace{0.5cm}\\
\includegraphics[width=0.95\columnwidth]{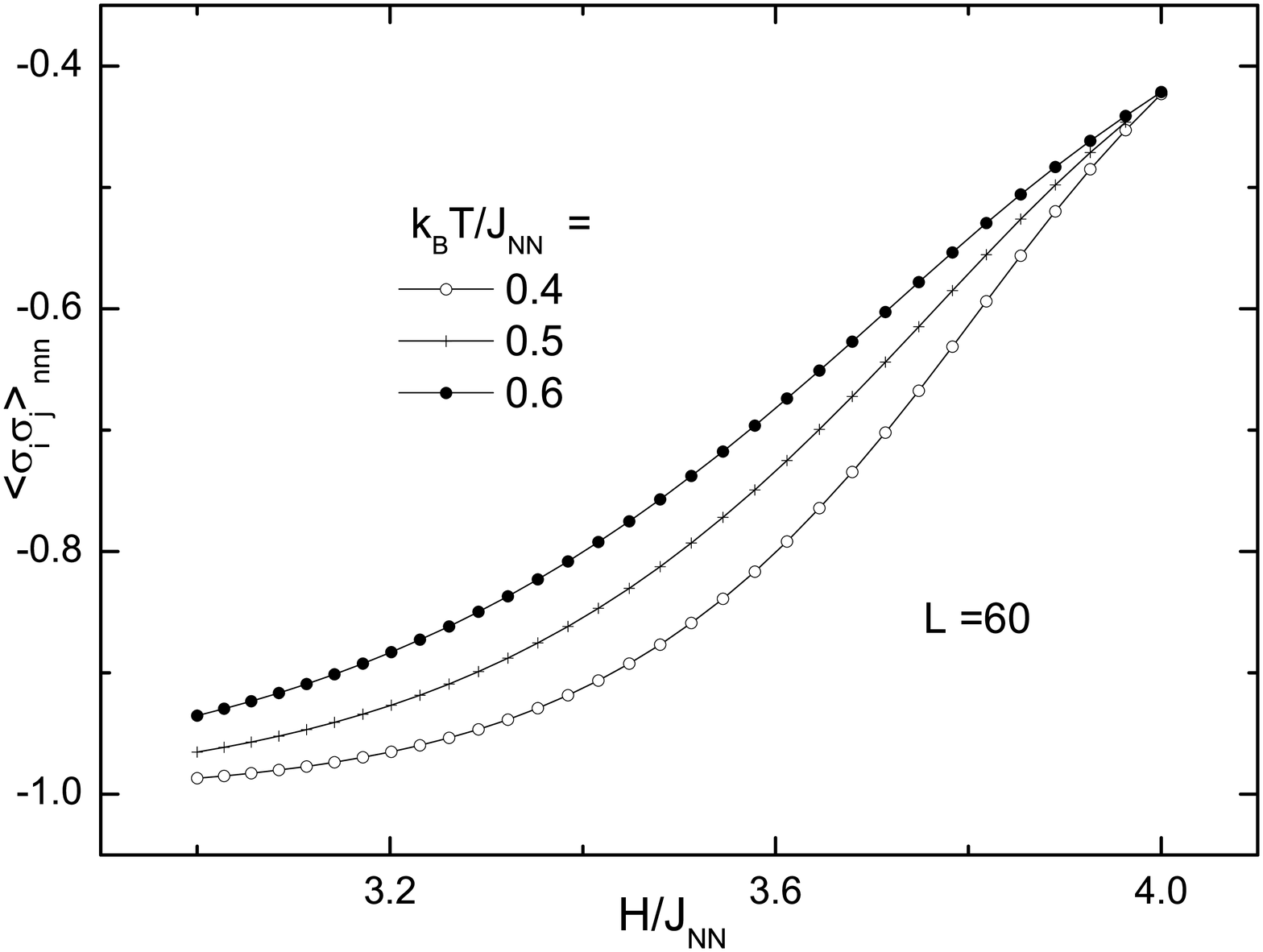}
\caption{Nearest- and next-nearest-neighbor correlation functions.
The field is varied for paths of 3 different temperatures:
$k_BT/J_{NN}=0.4,0.5$ and 0.6 across the short range ordering line.}
\label{f4}
\end{figure}

In order to confirm the above argument, we also calculated the NN
and NNN pair correlation function, that is $<\sigma_i\sigma_j>$, for
paths of different fields crossing this line. The correlation
function data are plotted in Fig.~\ref{f4}. The NN pair correlation
decreases from zero to a minimum and then increases to positive
values, while the NNN pair correlation increases monotonically from
$-1$.

\subsection{Critical behavior}
\begin{figure}
\includegraphics[width=0.95\columnwidth]{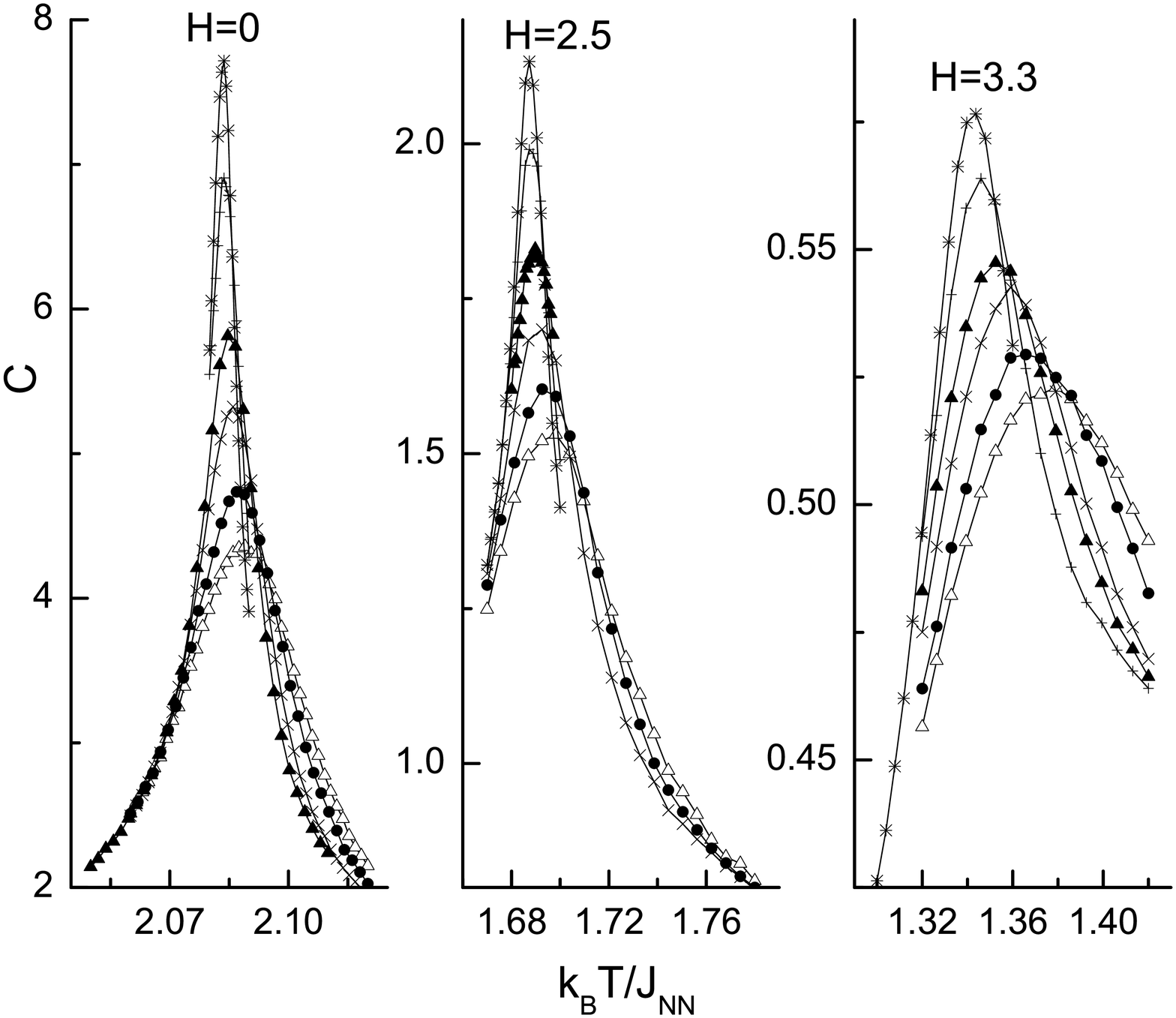}
\includegraphics[width=0.95\columnwidth]{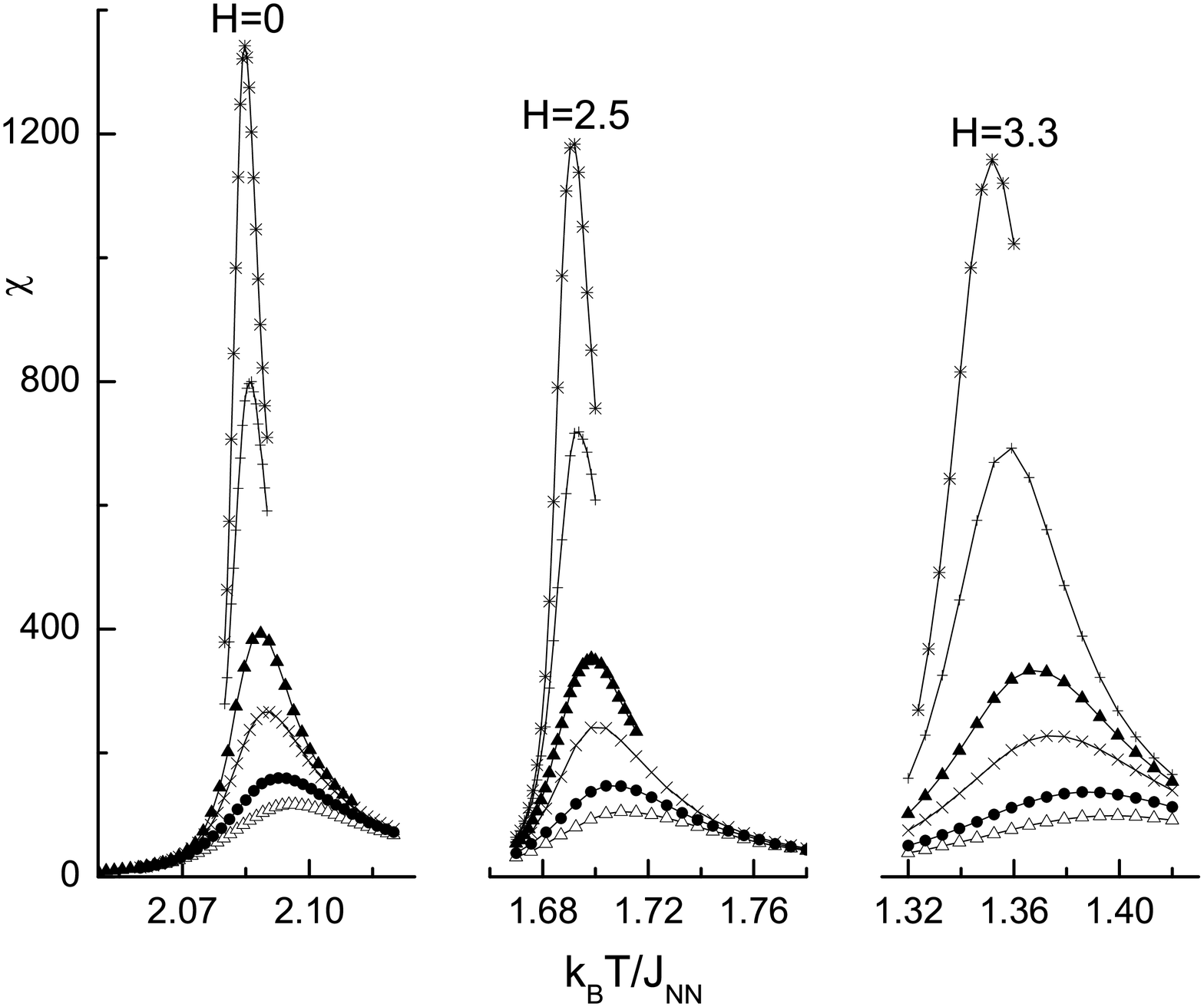}
\caption{Specific heat and susceptibility for 3 different fields
 across the phase boundary. Data are for:
L=100, $\vartriangle$; L=120, $\bullet$; L=160, $\times$; L=200
$\blacktriangle$; L=300 $+$; L=400 $\ast$.} \label{f5}
\end{figure}

The data for the specific heat and susceptibility for three different
values of $H$ are plotted in Fig.~\ref{f5}.

\begin{figure}
\includegraphics[width=0.9\columnwidth]{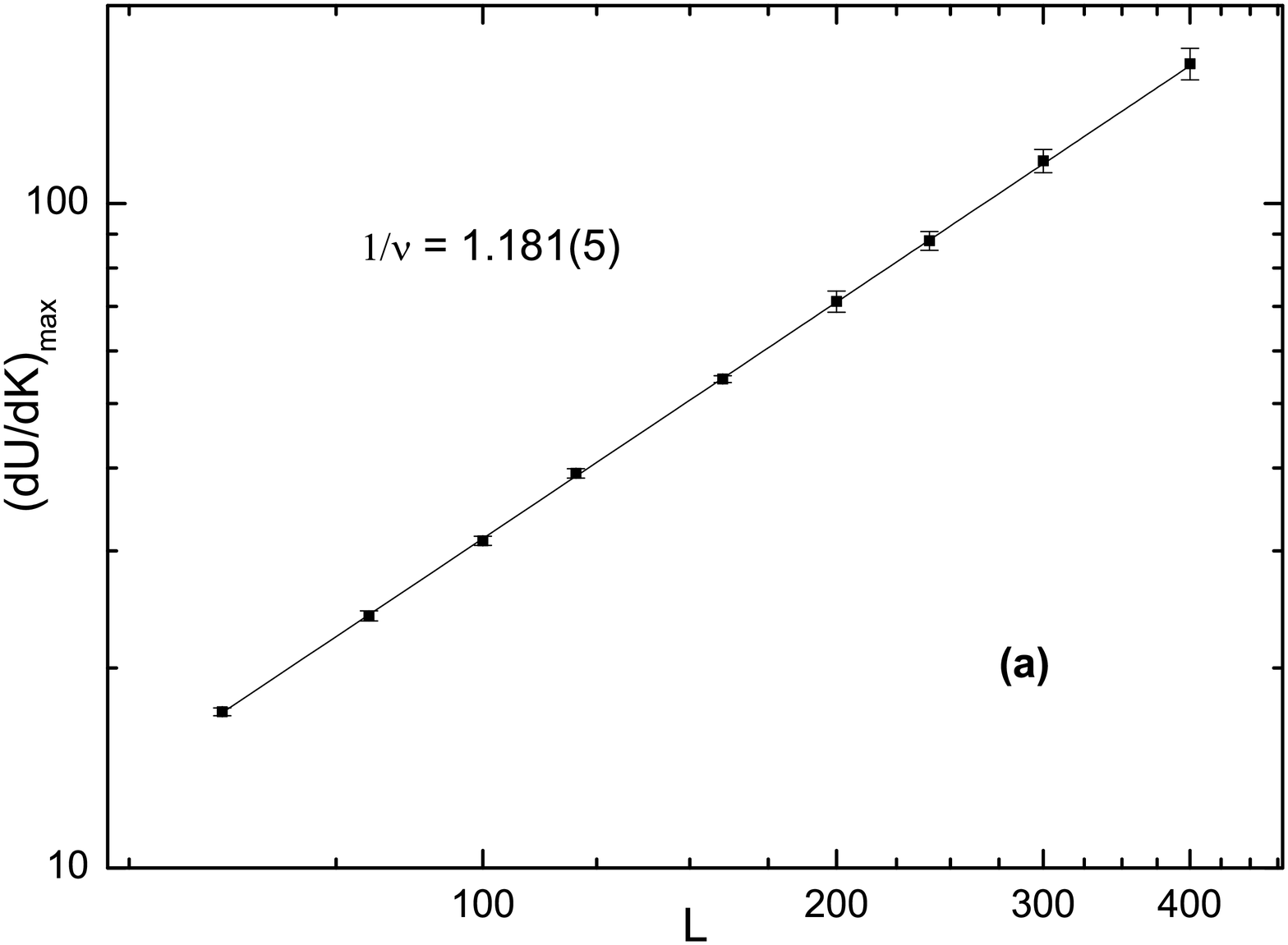}
\vspace{0.5cm}\\
\includegraphics[width=0.9\columnwidth]{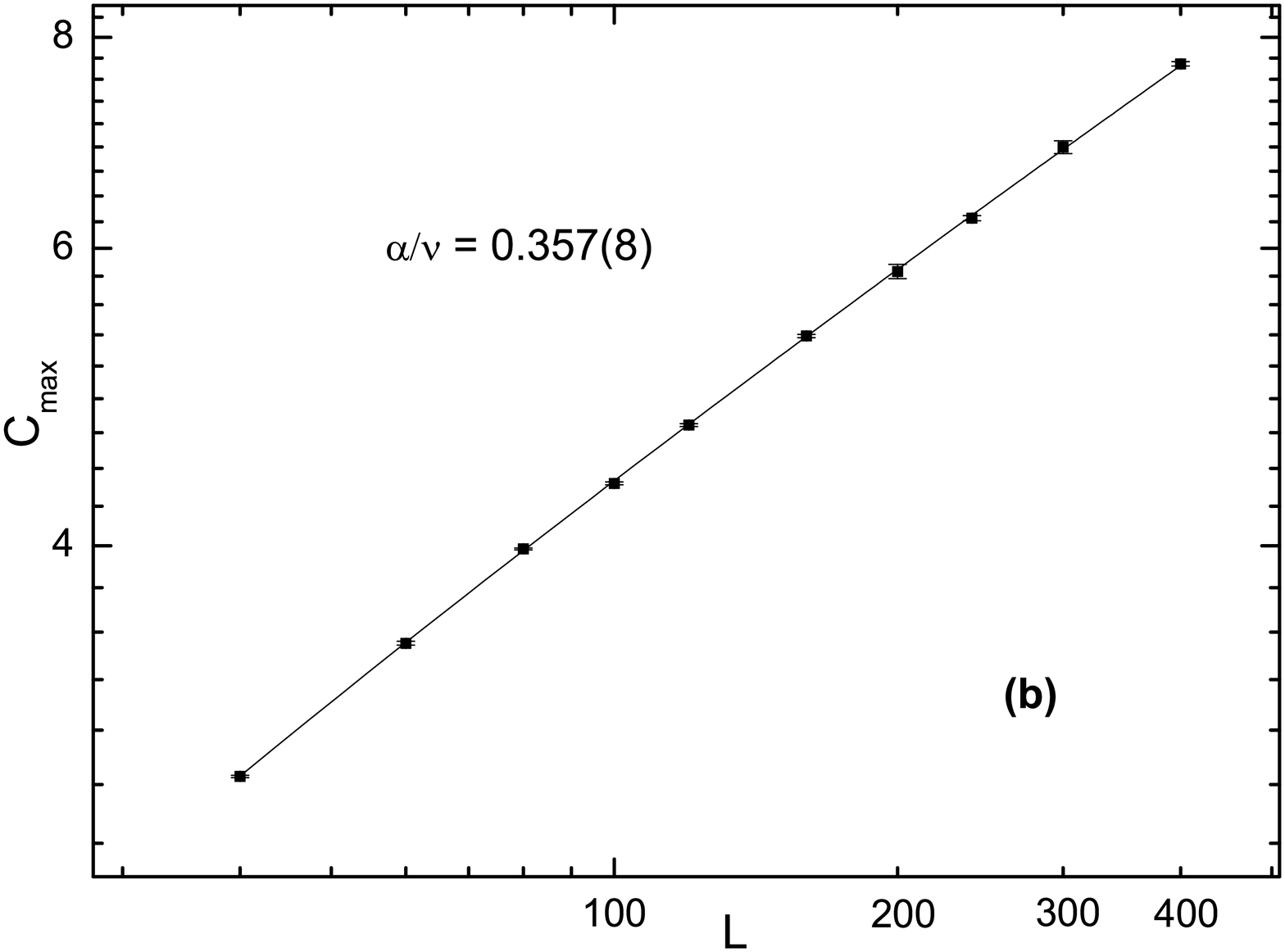}
\caption{Curve fits using the leading terms of equations \ref{eq1}
and \ref{eq2} for the maximum slopes of $\frac{dU}{dK}$ (a) and peak
values of specific heat (b), respectively.} \label{f6}
\end{figure}

Without the field, they
both show very sharp peaks, and from the magnitudes of the peak values of the specific
heat, as shown in Fig.\ref{f6}(b), we can obtain a rather accurate
estimate of the exponent ratio $\alpha/\nu=0.357(8)$, which is
obviously not zero. In Fig.\ref{f6}(a), we also show the curve-fit
for the maximum slope of $\frac{dU}{dK}$ for $H=0$, and extract
the exponent $\nu=0.847(4)$ directly.

Both values of $\alpha$ and
$\nu$ are quite consistent with the early estimates in
ref~\cite{Bin80}, and the value of $\alpha/\nu$ is different from
ref~\cite{Mal06}, in which the $1/L$ correction term was assumed up
to lattice size $L=160$.

The same procedure was repeated for $H/J_{NN}=2.5$ and $3.3$,
however, as shown in Fig.~\ref{f5}, the peaks of the specific
heat become increasingly rounded as the field increases, which makes it rather
difficult to get a direct estimate of the exponent $\alpha$.  Because of this it was necessary to obtain data for much larger lattice sizes, a task that was only possible with the use of GPU computing. In
fact, as the value of $\nu$ increases with the field, for
$H/J_{NN}=3.3$, according the hyper-scaling law $\alpha=2-d\nu$,
where $d=2$ is the dimension of the system, $\alpha$ should be
negative. Although the curve-fit is not stable, given the value of
$\alpha$ we can get a fit within error bars. Such continuous
increasing of the exponent $\nu$ up to values much greater than one
is actually consistent with the findings of an early transfer-matrix
study\cite{Ama83}.

\begin{figure}
\vspace{0.3cm}
\includegraphics[width=1.0\columnwidth]{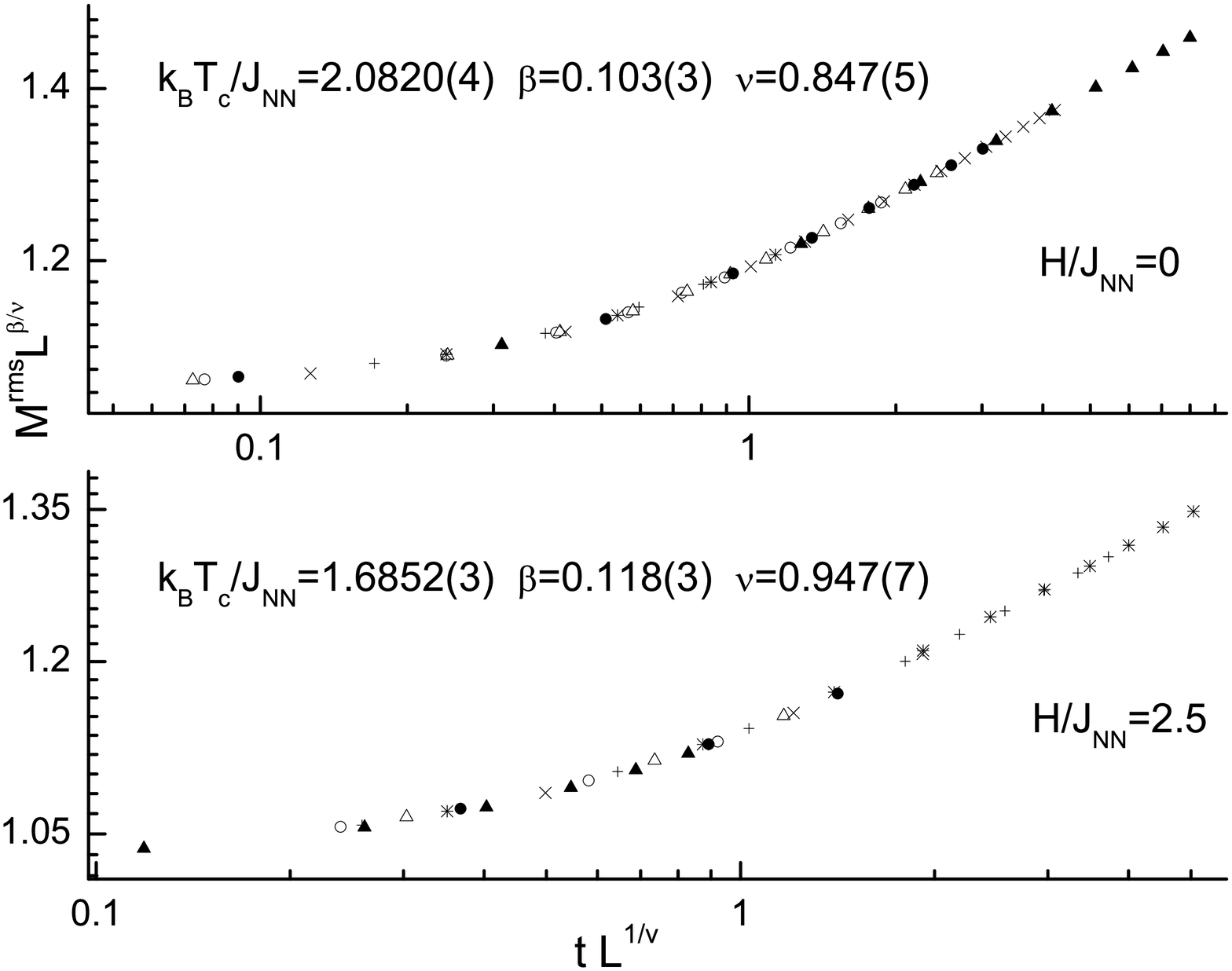}
\vspace{0.5cm}\\
\includegraphics[width=1.0\columnwidth]{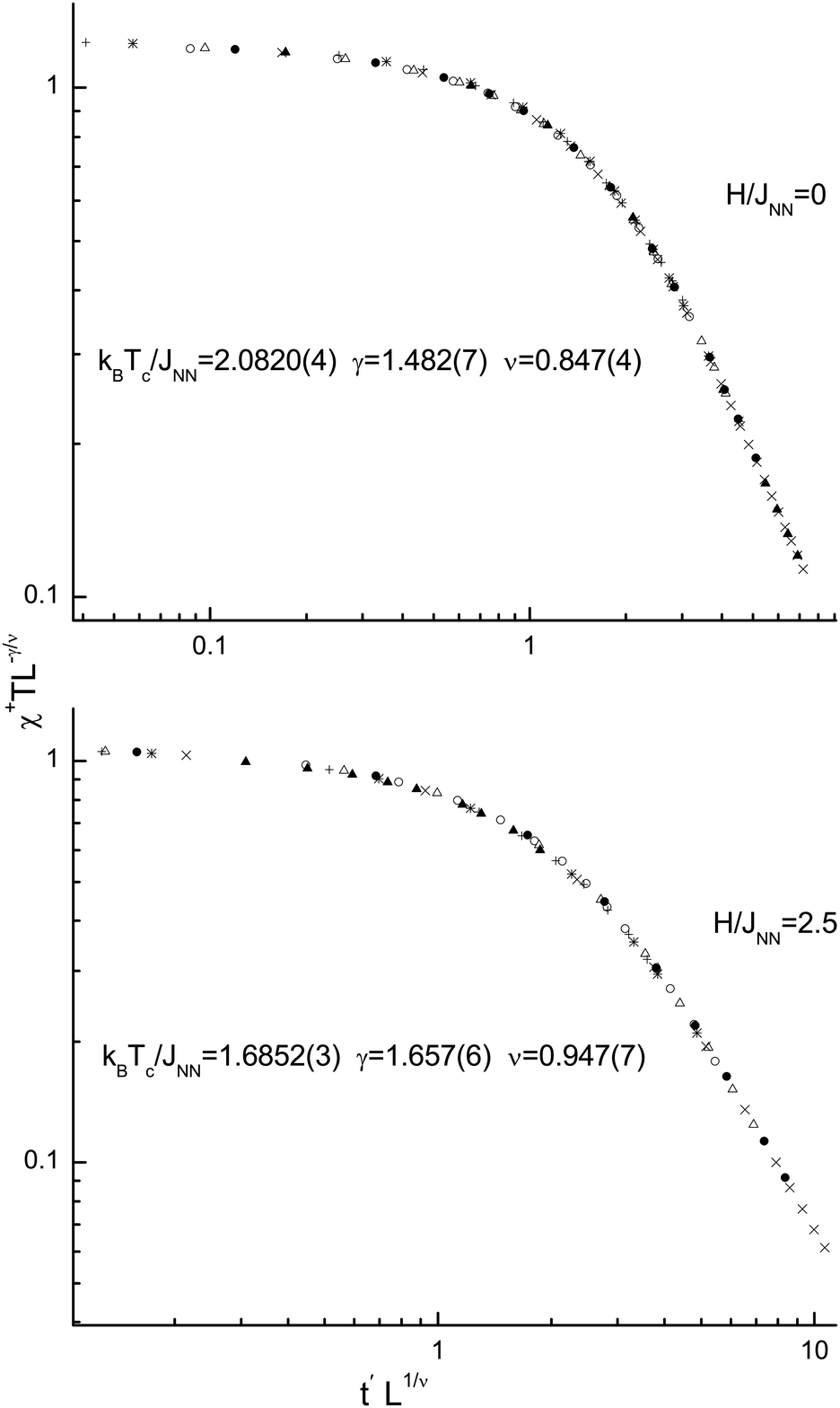}
 \caption{Finite size scaling data collapsing along
paths of constant $H/J_{NN}=0$ and $2.5$ for root-mean-square order
parameter and its ordering susceptibility, respectively.
$t^{\prime}=|1-\frac{T_c}{T}|$ and $t=(1-\frac{T}{Tc})$. Data are
for: L=80, $\circ$; L=100, $\vartriangle$; L=120, $\bullet$; L=160,
$\times$; L=200 $\blacktriangle$; L=300, $+$; L=400, $\ast$.}
\label{f7}
\end{figure}

To estimate the critical exponents $\beta$ and $\gamma$, we performed data
collapsing with a large range of lattice sizes for the order
parameter and its susceptibility. As shown in Fig.~\ref{f7}, the data in both finite size scaling plots
collapse very nicely onto single curves, and the ratio $\beta/\nu$ and $\gamma/\nu$
 agree with values of the 2D Ising universality class within error bars.

\begin{figure}
\vspace{0.3cm}
\includegraphics[width=1.0\columnwidth]{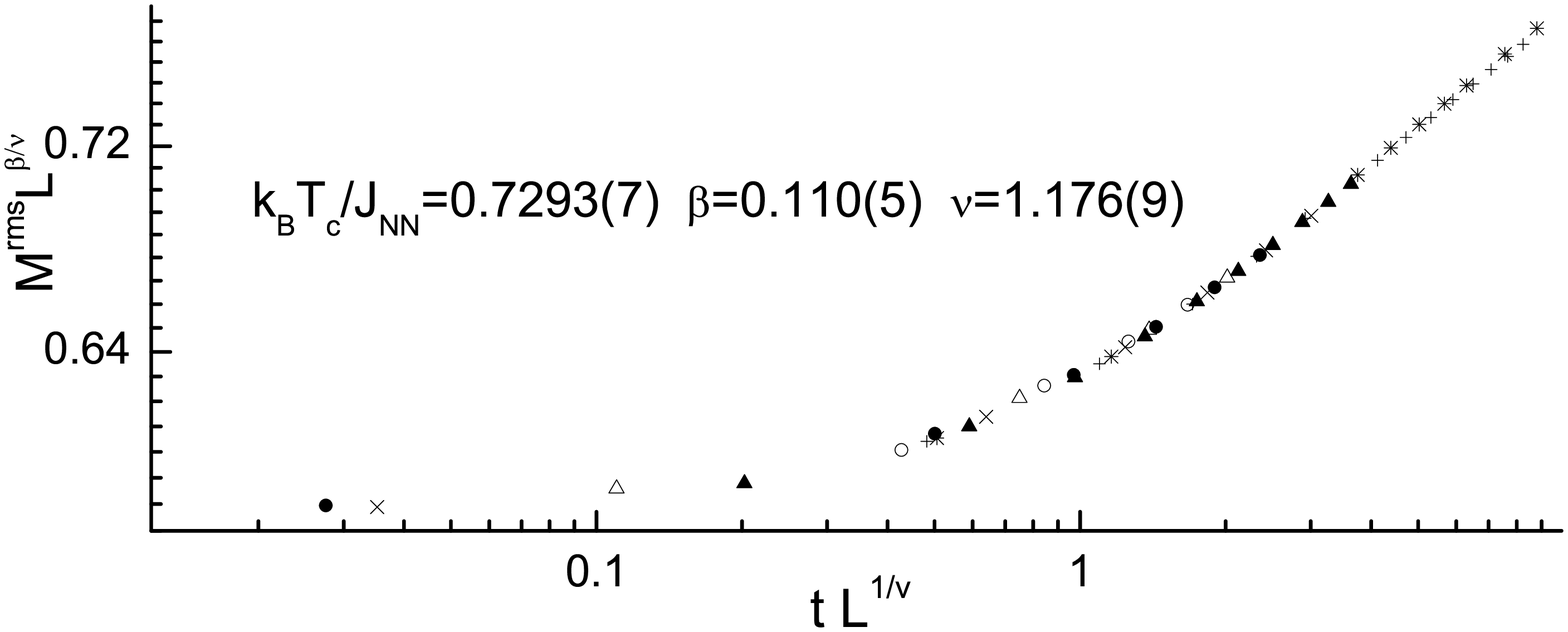}
\vspace{0.5cm}\\
\includegraphics[width=1.0\columnwidth]{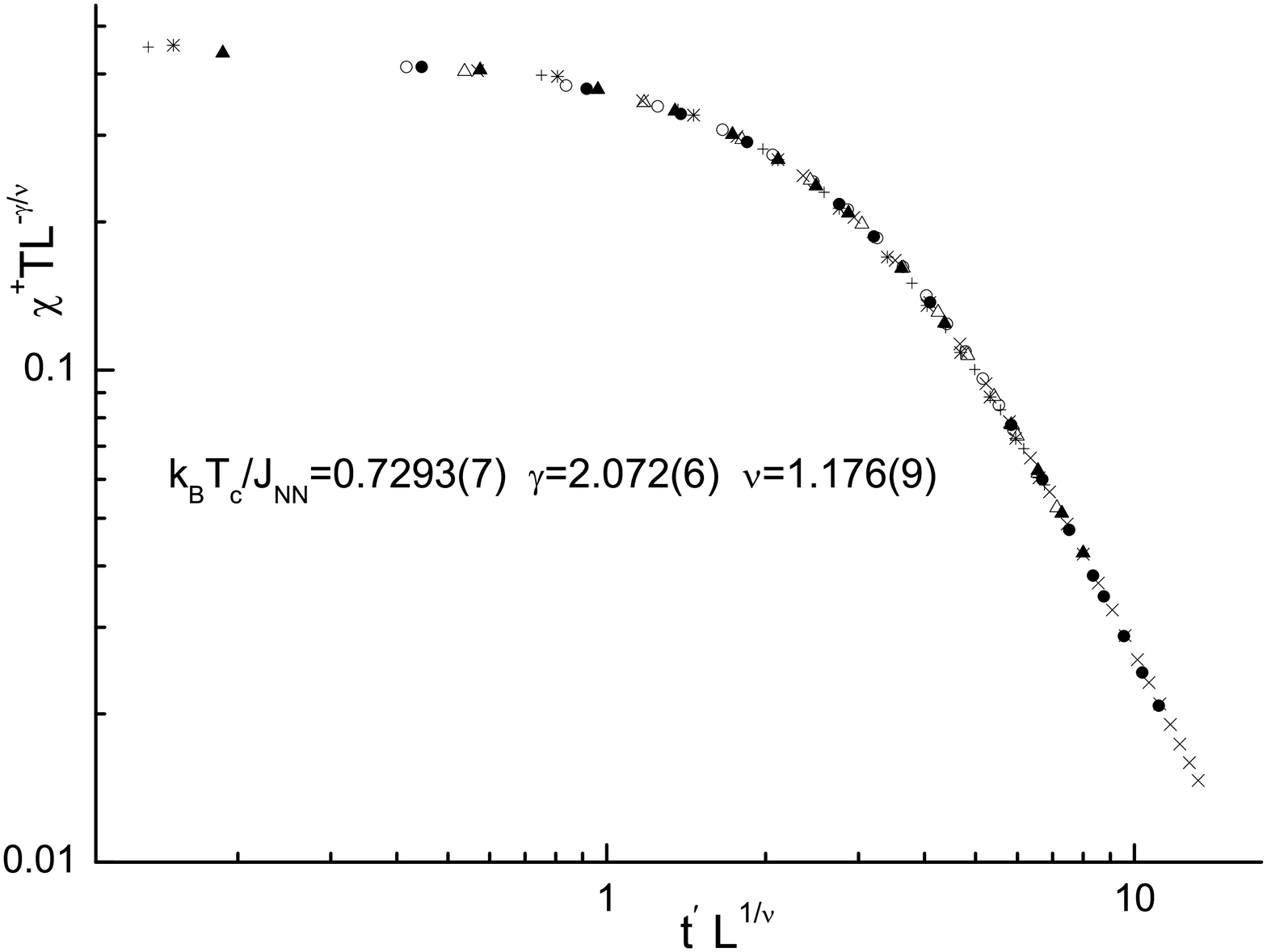}
\caption{Data collapsing along the path of constant $H/J_{NN}=6$ for
root-mean-square order parameter and its ordering susceptibility,
respectively. Data are for:  L=80, $\circ$; L=100, $\vartriangle$;
L=120, $\bullet$; L=160, $\times$; L=200 $\blacktriangle$; L=300,
$+$; L=400, $\ast$.} \label{f8}
\end{figure}

Hence, although the individual exponents are non-universal, Suzuki's
weak universality holds quite well. Another data collapsing along
the path of constant $H/J_{NN}=6$ across the phase boundary of the
row-shifted $2\times2$ state is shown in Fig.~\ref{f8}.  The quality
of the data collapsing is also excellent, and again, the exponents
are non-universal.  The estimate for $\beta/\nu$ is a bit low but
$\gamma/\nu$ agrees well with prediction of weak universality.

 In Table.~\ref{t1}, the critical points and exponents $\alpha,
\beta, \nu$ and $\gamma$ for several typical paths of constant H or T
across the phase boundary are listed.

\begin{table}[t]
\begin{center}
\begin{tabular}{cclllll} \hline \hline
   \multicolumn{1}{c}{$order$} &\multicolumn{1}{c}{$path $}   & \multicolumn{1}{c}{$T_c$ or $H_c$}
              & \multicolumn{1}{c}{$\alpha$}   & \multicolumn{1}{c}{$\beta$}
              & \multicolumn{1}{c}{$\gamma$}   & \multicolumn{1}{c}{$\nu$} \\ \hline
                                            &{H=0}&{$2.0820(4)$}   & {$0.302(7)$} & {$0.103(3)$} & {$1.482(7)$}  & {$0.847(4)$}    \\
         {$2\times1$}                       &{H=2.5}&{$1.6852(3)$}  & {$0.104(19)$} & {$0.118(3)$} & {$1.657(6)$}  & {$0.947(7)$}    \\
                                            &{H=3.3}&{$1.3335(6)$}   & {} & {$0.130(5)$} & {$1.930(6)$}  & {$1.102(8)$}
   \\ \hline
                                            &{H=6}&{$0.7293(7)$}   & {} & {$0.110(5)$} & {$2.072(6)$}  & {$1.176(9)$}    \\
   \raisebox{2.5ex}[0pt]{$2\times2$*}        &{T=0.5}&{$6.848(5)$}   & {} & {$0.126(4)$} & {$1.775(5)$}  & {$1.02(2)$}    \\
 \hline
\end{tabular}
\end{center}
\caption{ Values of critical point temperatures or magnetic fields and corresponding critical
exponents for several paths of constant temperature or field across the phase boundary
of the $(2\times1)$ and *row-shifted $(2\times2)$ ordered
phases.}\label{t1}
\end{table}

\subsection{Reentrance behavior}
Close to the region between the two ordered phases the correlation
length exponent $\nu$ turns out to be quite large, and
correspondingly the location of the critical points becomes very
difficult to determine. In addition, the specific heat curves look
"strange", see Fig.~\ref{f3}, because the exponent $\alpha/\nu$
would have a negative value with large magnitude, much larger
lattice size is needed to approach to the limiting peak value. Since
Suzuki's weak universality seems to hold along the transition line,
we fixed the values of $\beta/\nu=0.125$ and $\gamma/\nu=1.75$ for
the data collapsing analysis to get a better estimate of the
critical point and the exponent $\nu$.

\begin{figure}
\vspace{0.3cm}
\includegraphics[width=1.0\columnwidth]{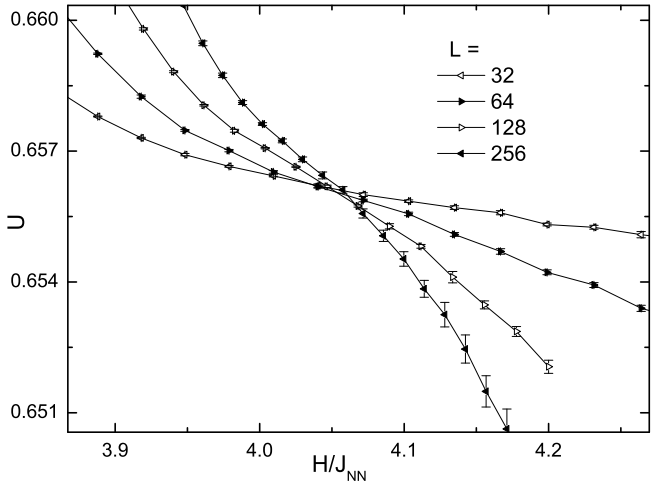}
\caption{Fourth order cumulant $U$ versus field $H$ along the path
of constant $k_BT/J_{NN}=0.7$ for lattice size $L=32,64,128,256$ .}
\label{f9}
\end{figure}

\begin{figure}
\vspace{0.3cm}
\includegraphics[width=1.0\columnwidth]{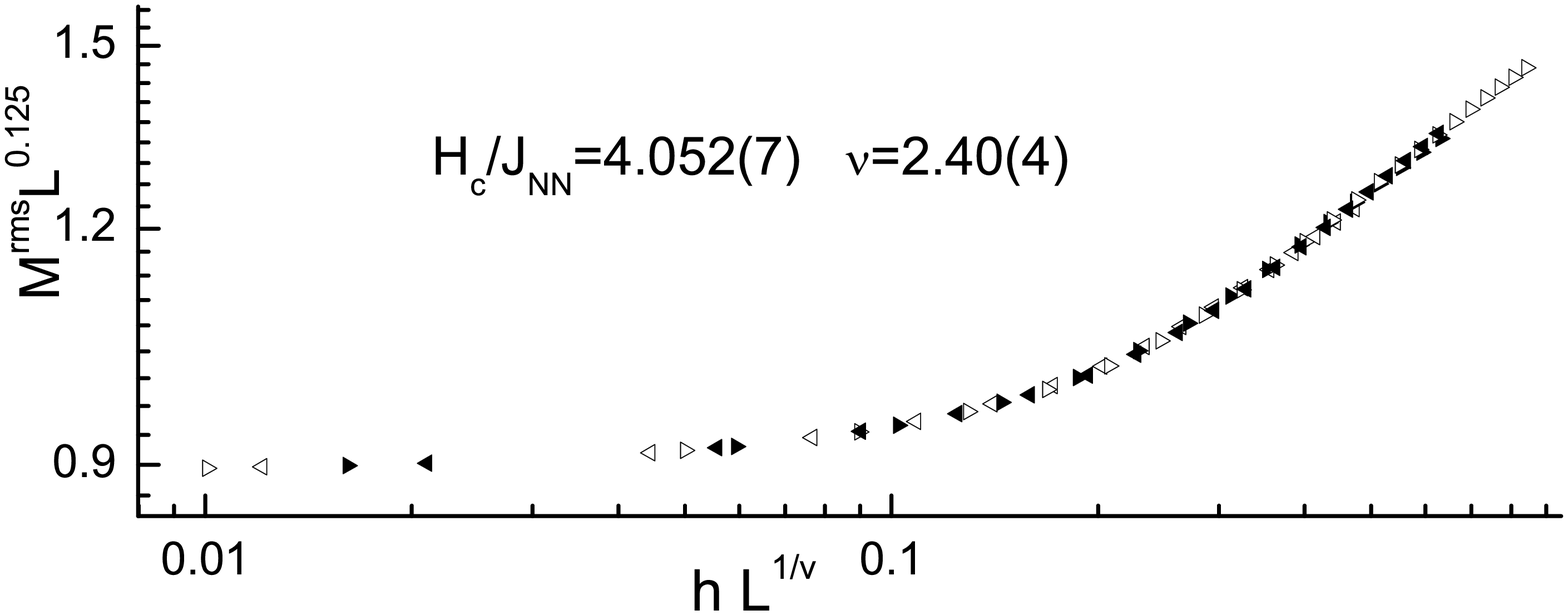}
\vspace{0.5cm}\\
\includegraphics[width=1.0\columnwidth]{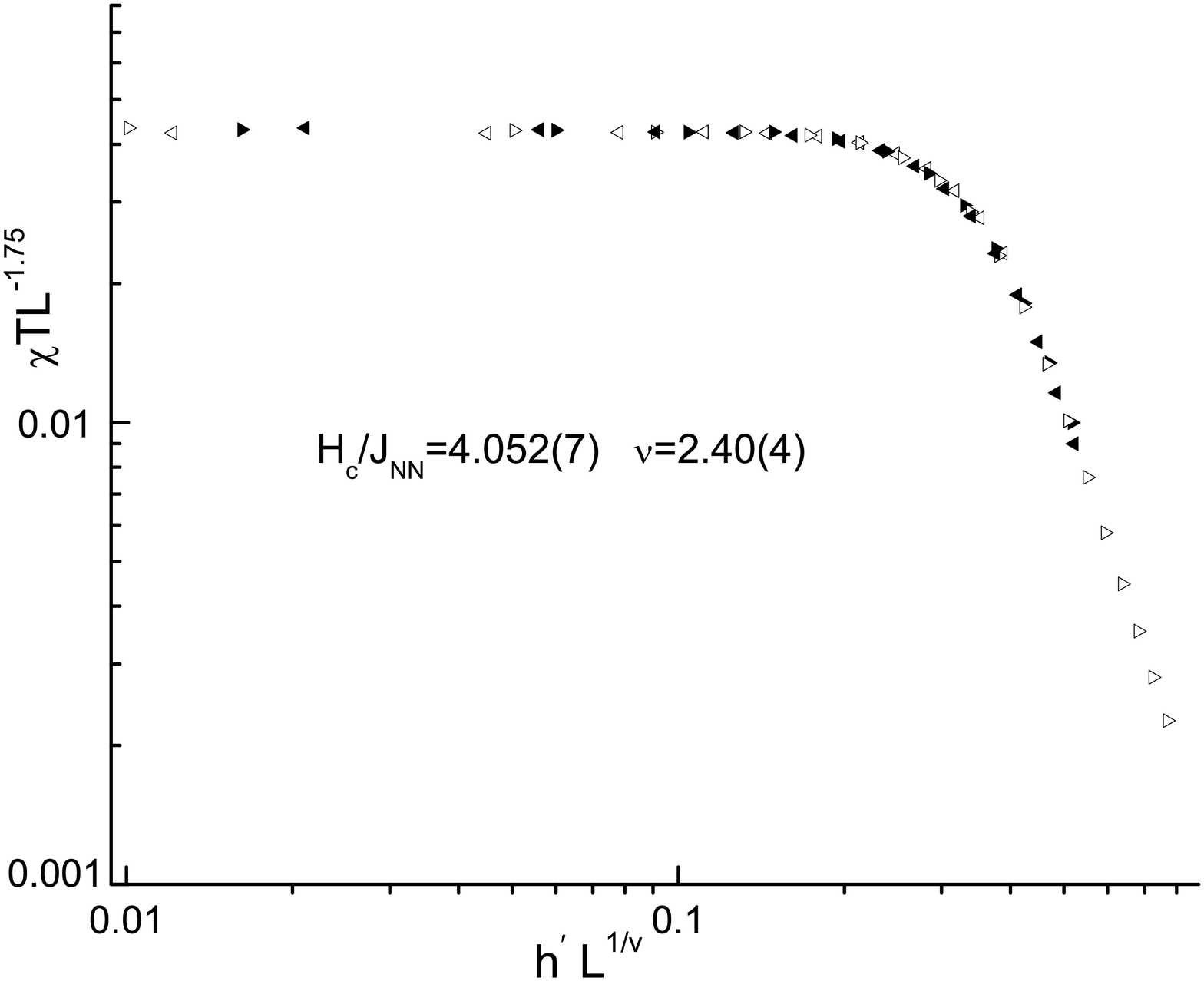}
 \caption{ Finite size scaling data collapsing along
the path of constant $k_BT/J_{NN}=0.7$ for root-mean-square order
parameter and its susceptibility, respectively.
$h^{\prime}=|1-\frac{H_c}{H}|$ and $h=(1-\frac{H}{Hc})$. Data are
for: L=32, $\triangleleft$; L=64, $\blacktriangleright $; L=128,
$\triangleright$; L=256, $\blacktriangleleft$.} \label{f10}
\end{figure}

As shown in Fig.~\ref{f9}, the crossing point of the fourth order
cumulant curves for a path of constant $k_BT/J_{NN}=0.7$ is slightly
above $H/J_{NN}=4$, and from the data collapsing analysis, see
Fig.~\ref{f10}, we obtained an estimate of the critical point to be
$H_c/J_{NN}=4.052(7)$.   Hence, we confirm the reentrant behavior of
the $(2\times1)$ transition line.

\begin{figure}
\vspace{0.3cm}
\includegraphics[width=1.0\columnwidth]{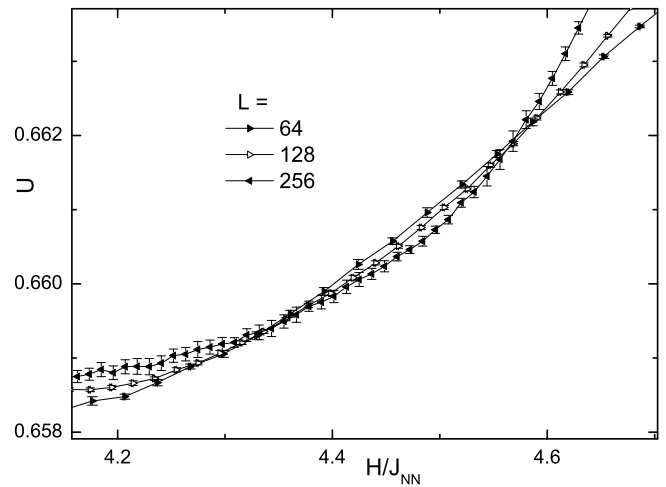}
\caption{Fourth order cumulant $U$ versus field $H$ along the path
of constant $k_BT/J_{NN}=0.45$ for lattice size $L=64,128,256$.}
\label{f11}
\end{figure}

For the paths of constant $k_BT/J_{NN}=0.45$, the curves of the
fourth order cumulant shows two crossing points and the finite size
effect is quite obvious. See Fig.~\ref{f11}.

\begin{figure}
\includegraphics[width=0.95\columnwidth]{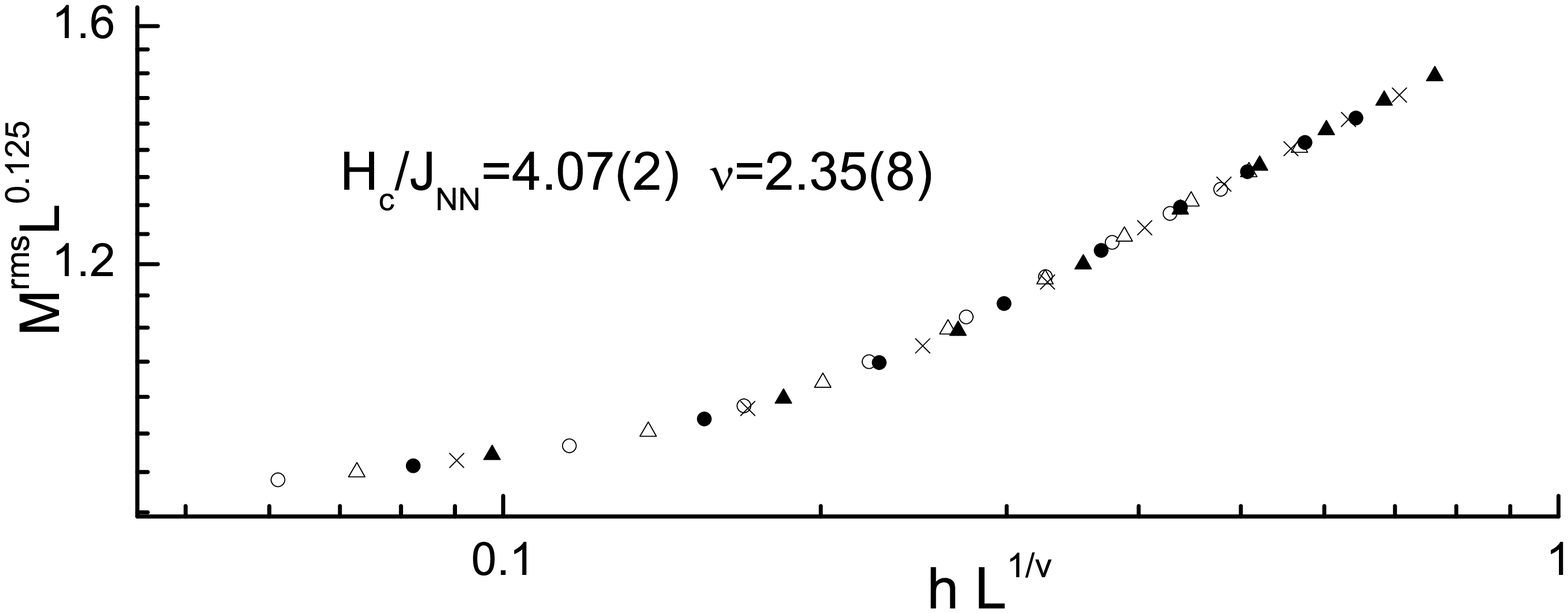}
\vspace{0.5cm}\\
\includegraphics[width=0.95\columnwidth]{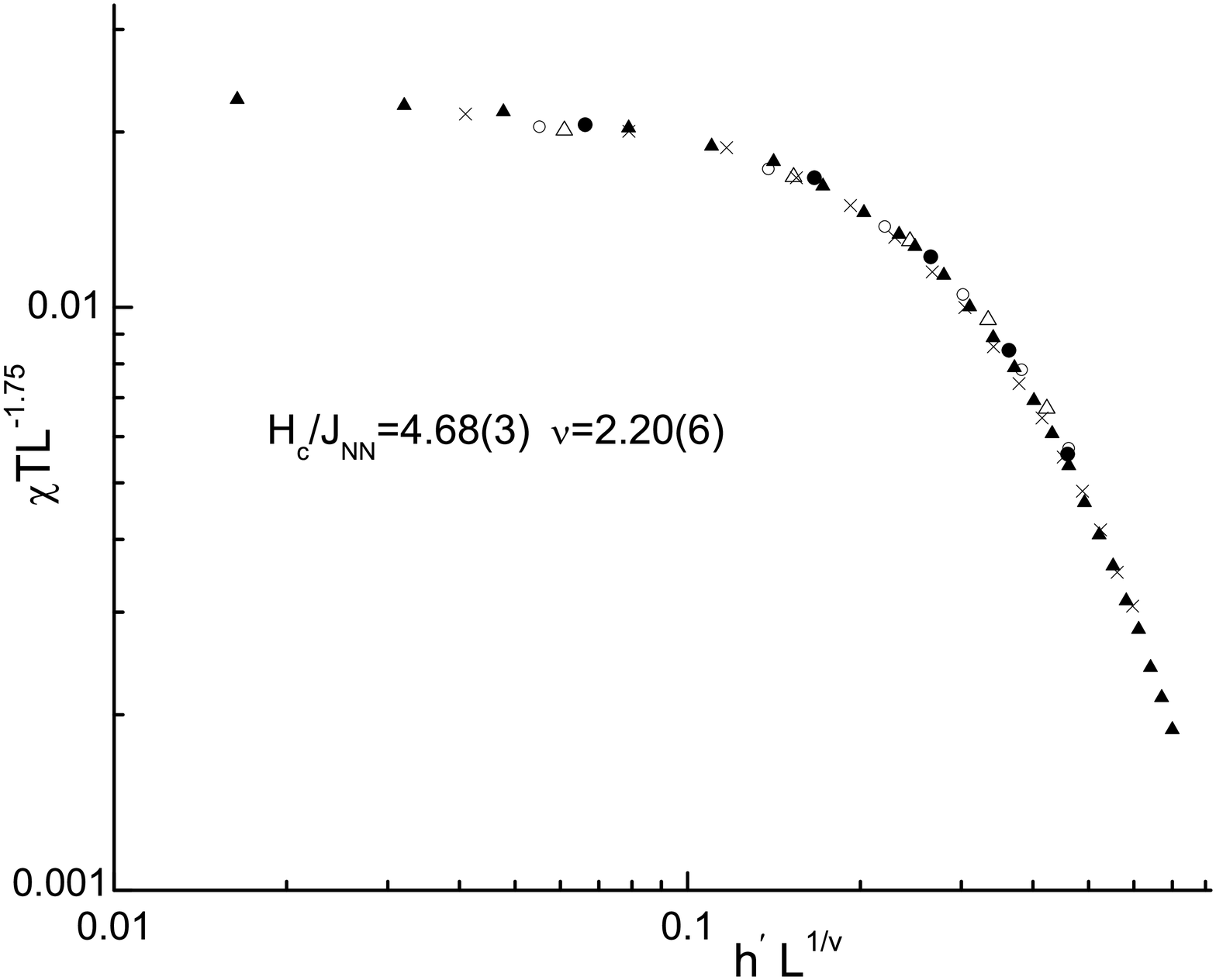}
\caption{Finite size scaling data collapsing along paths of constant
$k_BT/J_{NN}=0.5$ for root-mean-square order parameter and its
susceptibility, respectively. Data are for: L=80, $\circ$; L=100,
$\vartriangle$; L=120, $\bullet$; L=160, $\times$; L=200
$\blacktriangle$.} \label{f12}
\end{figure}

For the larger lattice size, the two crossing points move towards
lower fields but they do not approach each other.  Thus, a region of
disorder remains between the two different ordered states, even down
to quite low temperatures.  (If however, small lattices are used
with insufficient data  precision, it looks as though the curves for
different lattice sizes coincide.  Such behavior would indicate,
erroneously, the presence of an XY-like region.)  In Fig.~\ref{f12},
we show data collapsing fits along the path of constant
$k_BT/J_{NN}=0.5$ (and excellent data collapsing is also found along
the path of constant $k_BT/J_{NN}=0.3$), which confirms that there
is a disordered region in between the two ordered states.

We thus conclude that there is no XY-like region and that the two
phase boundaries probably only come together at a bicritical point
at $T=0$, although we cannot exclude the possibility of a bicritical
point at very low, but non-zero, temperature.  However, the data do
not yield any hint of such a bicritical point; but the lack of data
points for $T < 0.2$ in Fig.2 precludes us from making a definitive
statement about this issue.  (Moreover, the reentrant behavior of
the $(2\times1)$ phase makes it very difficult to study the approach
to $T=0$.) The variation of the critical exponents is consistent
with the changing magnetic field producing different effective
anisotropies which, in turn, is expected to yield non-universal
behavior\cite{Hu87}.  Due to the large values of $\nu$ near the
bicritical point (and correspondingly strongly negative values of
$\alpha$), we consider it also extremely unlikely that tricritical
points could be found along these transition lines as $T$ becomes
small.

\section{Conclusion}\label{s3}
We have carried out large-scale Monte Carlo simulations for the
square-lattice Ising model with repulsive (antiferromagnetic)
nearest- and next-nearest-neighbor interactions. From the finite
size scaling analysis, both transitions from $(2\times1)$ and
row-shifted $(2\times2)$ ordered states to disordered states turn
out to be continuous and non-universal. The reentrance behavior of
the $(2\times1)$ transition line is confirmed, and the proximity to
the transition line to the $(2\times2)$ state makes it difficult to
untangle the low temperature behavior unless quite large lattices
are used.  It was only possible to obtain the precise, large lattice
data needed though the use of GPU computing.  No evidence for
XY-like behavior is found, and we conclude that there is probably a
zero temperature bicritical point.  Although the exponent $\nu$
varies along the transition line, the exponent ratio $\beta/\nu$ and
$\gamma/\nu$ seem to agree with that of the 2D Ising universality
class.\\

{\bf Acknowledgements:} We would like to thank K. Binder and S.-H. Tsai for
illuminating discussions and comments and also T. Preis for introducing us to the use
of GPU's for our calculations.  Numerical computations
have been performed partly at the Research Computing Center of the
University of Georgia. This work was supported by NSF grant
DMR-0810223.

\clearpage


\end{document}